\documentclass[prl,twocolumn,showpacs]{revtex4}

\usepackage{epsfig}
\usepackage{graphicx}
\begin{document}

\title{Ferromagnetism in repulsive Fermi gases:\\ 
upper branch of Feshbach resonance versus hard spheres}
\author{Soon-Yong Chang, Mohit Randeria, and Nandini Trivedi}
\affiliation{Department of Physics, The Ohio State University, Columbus, OH 43210, USA}

\begin{abstract}
We use quantum Monte Carlo, including backflow corrections,
to investigate a two-component Fermi gas on the upper branch of a Feshbach resonance 
and contrast it with the hard sphere gas. We find that, in both cases, the Fermi liquid 
becomes unstable to ferromagnetism at a $k_F a$ smaller than  
the mean field result, where $k_F$ is the Fermi wavevector and $a$ the 
scattering length. Even though the total energies $E(k_F a)$ are similar in the 
two cases, their pair correlations and kinetic energies are 
completely different, reflecting the underlying potentials.
We discuss the extent to which our calculations shed light on recent experiments.
%The presence of a trap modifies the critical value of $k_F^0 a$ quantitatively. 
\end{abstract}

%71.27.+a       Strongly correlated electron systems; heavy fermions
%37.10.Jk       optical cooling/trapping of atoms
%67.85.-d			  ultracold gases
\maketitle

\noindent 
{\bf Introduction:} 
Ultracold atomic gases are emerging as a unique laboratory for testing quantum many-body Hamiltonians. A problem of fundamental importance is the ground state of two species of fermions
interacting via \emph{repulsive} interactions. The attractive case
is now well-understood and shows the BCS-BEC crossover~\cite{giorgini08} 
in the superfluid ground state.
The broken symmetry is already apparent within BCS mean field theory (MFT) with an
arbitrarily small attraction leading to a paired superfluid.
In contrast, we know much less about the repulsive case. The Landau Fermi liquid,
known to exist at weak repulsion~\cite{fetter}, can become unstable 
only beyond a critical value of the interaction\cite{stoner33}. Thus the phase transition is \emph{not} a weak coupling problem, and the validity of
MFT in the repulsive case is questionable.

An exciting new development is a recent experiment~\cite{ketterle09} which has been
interpreted as evidence for a ferromagnetic instability 
~\cite{stoner33,duine05,leblanc09,conduit09} in a ``repulsive'' Fermi gas of $^6Li$ atoms. 
A crucial point is that the interactions between the atoms 
are quite different from the textbook problem of hard-sphere interactions. 
In the experiment, the atoms are on the \emph{upper branch} of a Feshbach resonance
with a positive s-wave scattering length $a$.
The two-body ground state then is a molecule of size $a$. But in the 
upper branch, where the wave function is made up from scattering
states, the atoms feel an \emph{effective repulsion} characterized by $a > 0$, 
despite the fact that the underlying potential is attractive. 

The main question we examine in this Letter is the extent to which 
the many-body physics in the upper branch is similar to, or different from, 
that of a purely repulsive Fermi gas.
We use quantum Monte Carlo (QMC) to compute the energy,
chemical potential and pair distribution function of the two systems --
upper branch and repulsive -- to understand the instability of the
Fermi liquid to ferromagnetism. We believe that such a study of  
equilibrium properties is necessary, before one addresses 
non-equilibrium questions in the upper branch.
 
Before describing our results, we emphasize important ways in which our work differs
from previous studies, which focus on MFT of purely repulsive interactions. 
First, we carefully discuss what it means for
a many-body wavefunction to be on the upper branch, which is essential
to describe the experiments. Second, it is crucial to use QMC for this strong coupling problem.
For instance, QMC calculations~\cite{ceperley02} for the electron gas 
show that ferromagnetism sets in at a critical density nearly 3 orders of magnitude smaller 
than that predicted by Hartree-Fock MFT.
Finally, we include backflow corrections, which can have a nontrivial effect 
on the nodes of the many-body wavefunction, and thus on the ground state energy~\cite{kevin81}.
Not including backflow may lead to spurious ferromagnetic instabilities
in normal $^3$He \cite{ceperley-private}.

Our main results are that we find Ferromagnetic (FM) instabilities in both
the upper branch and the hard sphere Fermi gas. For small $k_F a > 0$, 
with $k_F$ the Fermi wavevector and $a$ the s-wave scattering length, both systems
are Landau Fermi liquids. The upper branch becomes unstable to 
a FM state at $k_F a = 0.89(2)$, independent of the details
of the interaction (in the zero-range limit). 
The critical $k_F a$ is similar for a purely repulsive interaction, 
but the result is \emph{non}-universal and depends on details the potential;
we will focus on hard spheres of diameter $a$.
In both cases the critical value is considerably smaller than the
Stoner MFT result $(k_F a)_{\rm MFT} = \pi/2$ \cite{stoner33}.
Despite similar values of the critical interaction, the behavior of the kinetic
energy and the two-body correlations are \emph{qualitatively different}
for the upper branch and hard spheres.
We also discuss the harmonically trapped gas using 
the local density approximation (LDA). We conclude with a brief comparison of our
results with experiments \cite{ketterle09}.  We find that 
some aspects of the experiment cannot be understood within our equilibrium theory.

%\begin{widetext}
\begin{figure*}[]
\includegraphics[width=5.5cm,clip]{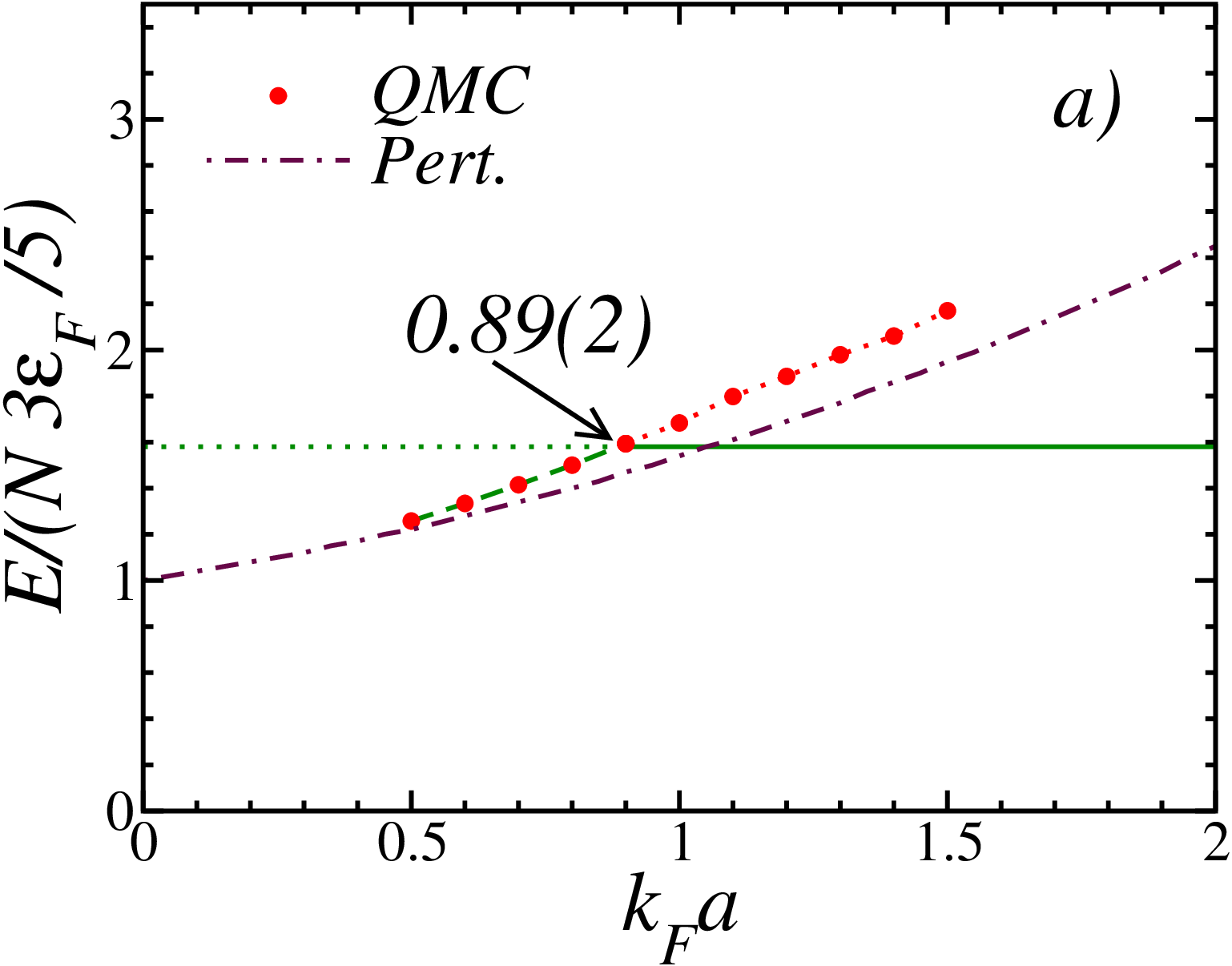}
\includegraphics[width=5.5cm,clip]{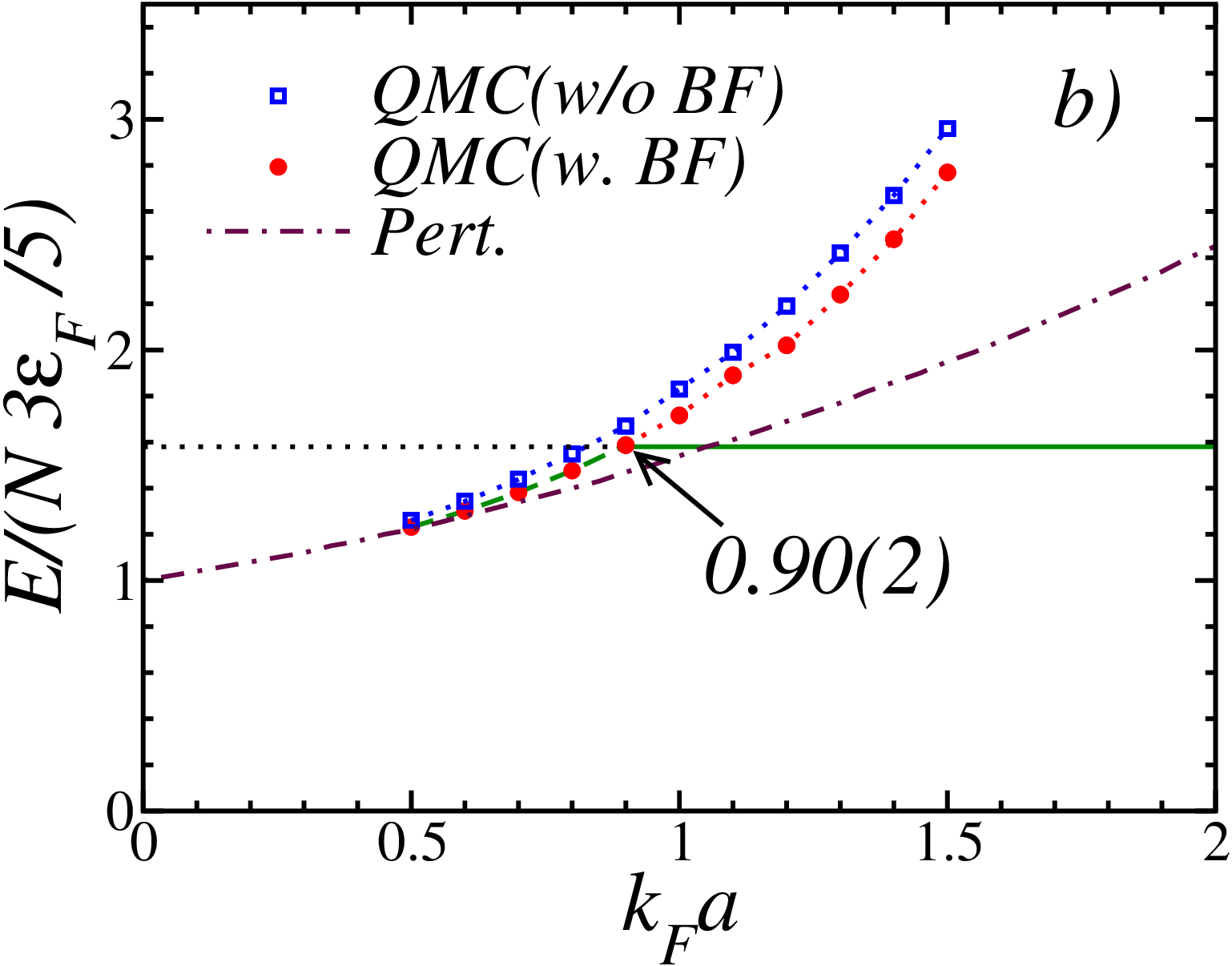}
\includegraphics[width=5.5cm,clip]{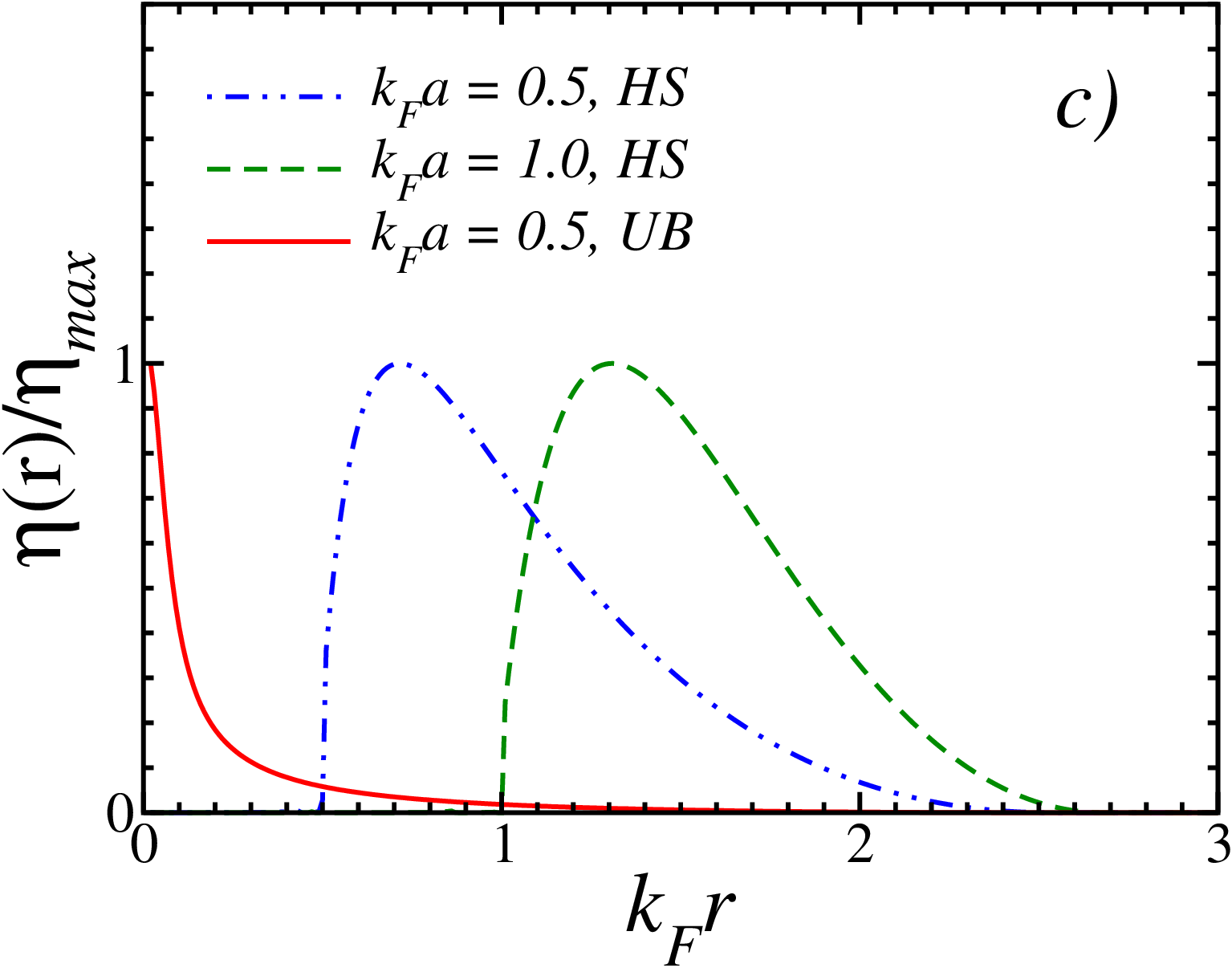}
\caption{(Color online) QMC energy per particle for the Fermi liquid state 
(for $N_\uparrow = N_\downarrow = 33$ particles)
as a function of $k_F a$ for (a) the upper branch and for (b) 
hard spheres, compared with the perturbative result Eq.~(\ref{pert-thy}).
Panel (b) shows QMC with and without backflow (BF) corrections. Backflow
does not have a significant effect on the upper branch results in (a). 
There is a transition to a ferromagnetic state
when the QMC energy crosses that of the fully polarized gas (horizontal lines). 
(c) The function $\eta(r)$ describing BF correlation, discussed in the text,
for the upper branch (UB) and hard sphere (HS) gas.}
\label{energy}
\end{figure*} 
%\end{widetext}

\noindent{\bf Model:}
We consider a gas of $N = (N_\uparrow + N_\downarrow)$ fermions of mass $m$ with two species, denoted by ``spin'' $\uparrow$ and $\downarrow$, which interact via a potential $V(r)$. The Hamiltonian is
\begin{equation} 
H=\sum_{i\sigma} { { {\bf p}_{i\sigma}^2} \over{2m}}  
+ {1\over 2}\sum_ {{i},{j}}V(r_{ij})
\label{eqn_hamiltonian}
\end{equation}
with $r_{ij} = \vert{\bf r}_{i\uparrow}-{\bf r}_{j\downarrow}\vert$.
For the QMC calculations we consider a cubic box with periodic boundary conditions. 
(At the end, we also discuss trap effects within LDA).
We measure lengths in units of $k_F^{-1}$, where $k_F=(3\pi^2 n)^{1/3}$ 
for a free Fermi gas of density $n$. 
We measure energies in units of $\epsilon_{FG}= 3\epsilon_{F}/5$ where 
$\epsilon_F=k_F^2/2m$ (with $\hbar=1$). 
 
We consider two forms of the interaction potential. 
For the repulsive case, we use $V(r)=V_0 > 0$ for $r<R$ and zero elsewhere.
In the hard sphere limit $V_0\rightarrow \infty$ and the
diameter of the sphere $R = a$ scattering length.
For the attractive case, we use
$V(r) = - (8/mR^2) V_0 / {\rm cosh}^2(2 r/R)$,
extensively used in BCS-BEC crossover studies \cite{carlson03,chang04}. 
We choose the range such that $k_FR \ll 0.1$, so that we
obtain universal results independent of the detailed
form of $V(r)$. We use $V_0$ to tune the
scattering length $a > 0$, such that $V(r)$ has a single bound state.
We will focus on scattering states to construct the upper branch wavefunction. 

\noindent
{\bf QMC Results:}
The many-body wave function for a (paramagnetic) Fermi liquid is of the Jastrow-Slater form
\begin{equation}
\Psi=\prod_{{i\uparrow},{j\downarrow}}f(r_{ij}) \Phi_{{FG}\uparrow} \Phi_{{FG}\downarrow}.
\label{eqn_wf}
\end{equation}
The Slater determinants $\Phi_{{FG}\sigma}$'s 
are constructed from plane waves,
and the symmetric Jastrow function $f(r)$ accounts for interactions.  

We now argue that the upper branch Jastrow factor must be qualitatively different from
the $f(r) \geq 0$ used for the purely repulsive case.
To see this, consider using a conventional nodeless $f(r)$ for the attractive Fermi gas. 
This state is a \emph{normal} Fermi liquid, and thus orthogonal (for large $N$) 
to the \emph{superfluid} ground state \cite{giorgini08,carlson03} 
of the BCS-BEC crossover for all $k_F a$. 
However, the energy per particle in this state
is always \emph{lower} than the free gas $3\epsilon_{F}/5$,
which means that the fermions do feel an attraction.
In other words, this normal wave function necessarily has some pairing
(bound state-like) correlations, and is therefore \emph{not} on the upper branch.

A necessary condition for a many-body state to be on the upper branch is that 
its energy per particle must be greater than $3\epsilon_{F}/5$. 
We must ensure that every pair of particles feels an effective repulsion.
We achieve this by introducing a node in the Jastrow $f(r)$. 
To determine $f(r)$, we use the lowest-order constrained
variational (LOCV) method~\cite{vijay1}, which is
well known in nuclear physics and has also been used for strongly 
interacting quantum gases~\cite{chang04,cowell02}.
The LOCV equation has an upper-branch solution~\cite{locv-footnote}
$f(r)$ with a node, whose location tracks the scattering length at 
small $a$ [i.e., $f(r)\sim (1 - a/r)$] but then saturates at large $a$.  

We use QMC to calculate the energy for the
upper branch [Fig.~\ref{energy}(a)] and the hard sphere Fermi gas [Fig.~\ref{energy}(b)]
with $N_\uparrow = N_\downarrow$.
For small $k_F a$, both results agree with the
well-known perturbative result \cite{fetter}  
\begin{equation}
\frac{E}{N\epsilon_F} = \frac{3}{5} + \frac{2}{3\pi} k_F a + \frac{4}{35\pi^2}(11-2\ln2) (k_F a)^2 
+ \ldots
\label{pert-thy}
\end{equation}
We note that Eq.~(\ref{pert-thy}) should be taken seriously \emph{only for} 
$k_F a \ll 1$; the third order term is known to be \emph{non}-universal,
and depends on the detailed shape of the potential and on the p-wave scattering 
channel~\cite{fetter}.

A sufficient criterion for ferromagnetism (FM) is that
the energy of the paramagnetic Fermi liquid state exceed that
of the fully polarized state $\epsilon_{FG}^P/(3\epsilon_F/5)=2^{2/3} \simeq 1.58$.
It is instructive to begin with simple analytical approximations 
(even though these involve using Eq.~(\ref{pert-thy}) beyond its domain of validity!)
The simplest approximation is to just keep the first term
in (\ref{pert-thy}), the mean field Hartree shift. We find that this energy
crosses that of the fully polarized $\epsilon_{FG}^P$
at $k_F a = (2^{2/3} - 1)9\pi/10 \simeq 1.66$. This is slightly larger than the
Stoner estimate of $\pi/2$, but still below the hard sphere solidification limit 
$k_F a = (9\pi/4)^{1/3} \simeq 1.92$. 
Including the second order term in (\ref{pert-thy})
increases the energy of the paramagnetic Fermi-liquid solution, and 
thus FM sets in closer to $k_F a \simeq 1$.

The QMC energy for both the upper branch and hard spheres
implies a FM ground state for $k_F a \gtrsim 0.9$.
We next address backflow to see how it affects
our conclusion.

\noindent 
{\bf Backflow:}
It is very important to include backflow which,
as noted above, makes nontrivial modifications to the
nodal surfaces and can lead to large quantitative 
effects~\cite{kevin81} in the ground state energy. 
Backflow modifies the single-particle plane wave orbitals 
$\phi_{\bf k}({\bf r}_{i\sigma}) = \exp\left[i{\bf k}\cdot{\bf r}_{i\sigma}\right]$
used to construct the Slater determinants in Eq.~(\ref{eqn_wf}) via the 
replacement
${\bf r}_{i\sigma} \rightarrow {\bf r}_{i\sigma} + \sum_{j} \eta(r_{ij}){\bf r}_{ij}$,
where $j$ labels particles of the opposite spin $\bar{\sigma}$.

The optimal form of the backflow function $\eta(r)$ must be determined for each problem;
it is known to be very different for the $^4$He roton~\cite{feynman56}, 
for normal $^3$He \cite{vijay2}, and for the electron gas \cite{kwon93}. 
Insight into the form of $\eta(r)$ for $^3$He 
came from analyzing the problem of a $^3$He impurity in $^4$He \cite{vijay2}.
Following the same logic, we consider a single 
spin-down impurity in a spin-up Fermi sea. 
Omitting the details of our analysis (which will be reported elsewhere),
we find the $\eta(r)$'s shown in Fig.~\ref{energy}(c).

For the hard sphere gas the optimal $\eta(r)$ vanishes inside the
hard-core diameter and has a single peak just beyond it,
qualitatively similar to the case of $^3$He \cite{vijay2,kevin81}.
As in $^3$He, we approximate the form of $\eta(r)$ by a Gaussian
whose parameters we optimize. We use QMC to compute the 
energy of hard spheres using Eq.~(\ref{eqn_wf}) with 
a nodeless Jastrow times backflow-corrected Slater determinants.
We find that backflow leads to a significant lowering of energy
[see Fig.~\ref{energy}(b)] that becomes more pronounced with increasing $k_F a$.
For example, there is a $5.5\%$ reduction in energy at $k_F a = 1$.

For the upper branch, we find that the 
form of the optimal $\eta(r)$ is qualitatively different; see Fig.~\ref{energy}(c).
It is nonzero at the origin and decreases monotonically,
with a power-law decay at large $r$. Further, $\eta(r)$ changes very
little with $k_F a$ compared with the hard sphere case. 
The form of upper branch $\eta(r)$ is similar to 
systems with a soft-core, long-range repulsion, like the electron gas~\cite{kwon93}.
We use QMC to calculate the energy of the upper branch state 
(\ref{eqn_wf}), with a Jastrow with a single node, times 
backflow-corrected Slater determinants. 
In this case the reduction in energy is small 
[Fig.~\ref{energy}(a)] and falls within our statistical error of 
$\lesssim 1\%$.

We thus find that backflow is important for hard spheres
when $k_F^{-1}$ is comparable to the hard-core diameter $R=a$. 
On the other hand, backflow effects are small for the upper branch,
where $k_F^{-1} \gg R$, the range. 

\noindent 
{\bf Observables:}
For both the upper branch and for hard spheres,
we conclude that ferromagnetism is energetically favorable,
based on the crossing of energies of the paramagnetic Fermi liquid
and the fully polarized FM; see Fig.~\ref{energy}(a,b). For the upper branch,
we find that FM state is stable for $k_F a \geq 0.89(2)$.
The order of the transition requires a careful finite-size scaling
analysis in the vicinity of the phase transition, beyond 
the scope of our present investigation. 

\begin{figure}[]
\includegraphics[width=5.5cm,clip]{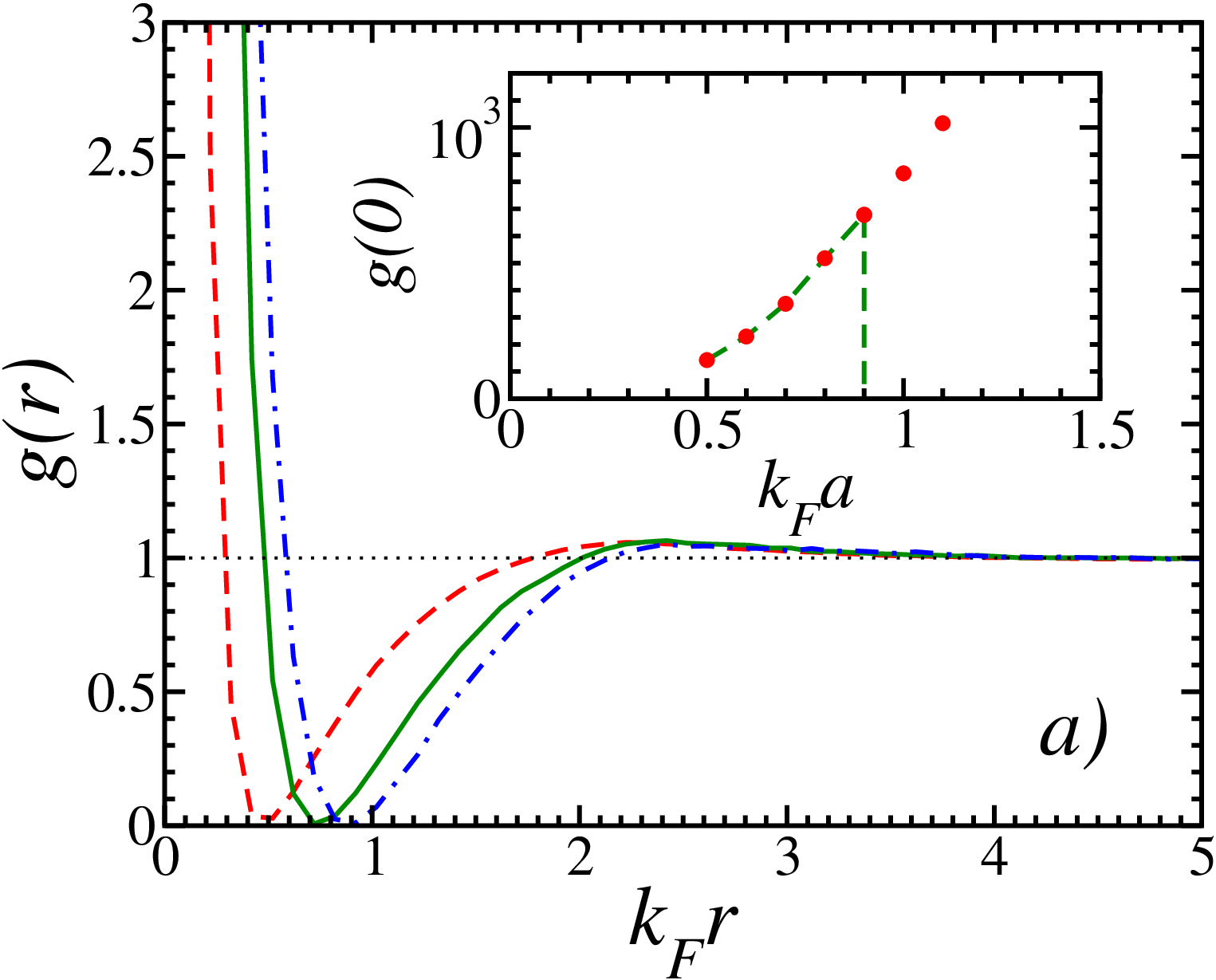}
\includegraphics[width=5.5cm,clip]{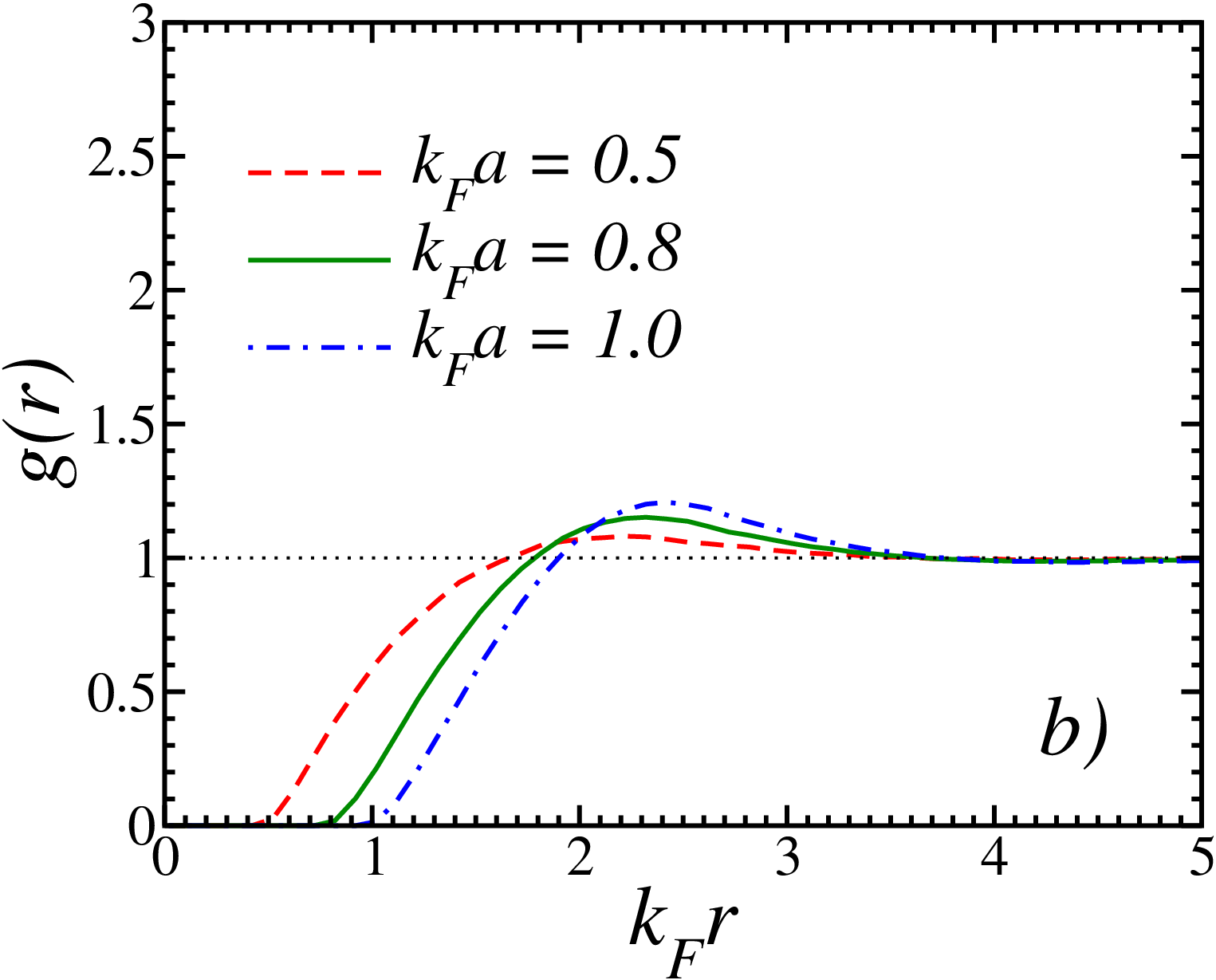}
\caption{(Color online)
Pair distribution function $g(r) \equiv g_{\uparrow\downarrow}(r)$ 
for (a) the upper branch and (b) the hard sphere gas for various values of $k_F a$.
Inset in panel (a) shows $g(r=0)$ as a function of $k_F a$.
}
\label{g2}
\end{figure}

Although the total energies in the Fermi liquid phases in the upper branch and hard spheres
are similar, the potential $\langle V \rangle$
and kinetic energy $\langle K \rangle$ are very different in the two cases.
To understand this, it is illuminating to look at the pair distribution function 
$g_{\uparrow\downarrow}(r)$, denoted by $g(r)$ for simplicity.
In the hard sphere case [Fig.~\ref{g2}(b)], $g(r)$ vanishes inside the 
hard-core and goes to unity at large separation.
The potential energy $\langle V\rangle \sim \int d^3{\bf r} g(r) V(r)$
then vanishes identically and the total energy [Fig.~\ref{energy}(b)] 
in the hard sphere case is entirely kinetic. 

\begin{figure}[]
\includegraphics[width=6.0cm,clip]{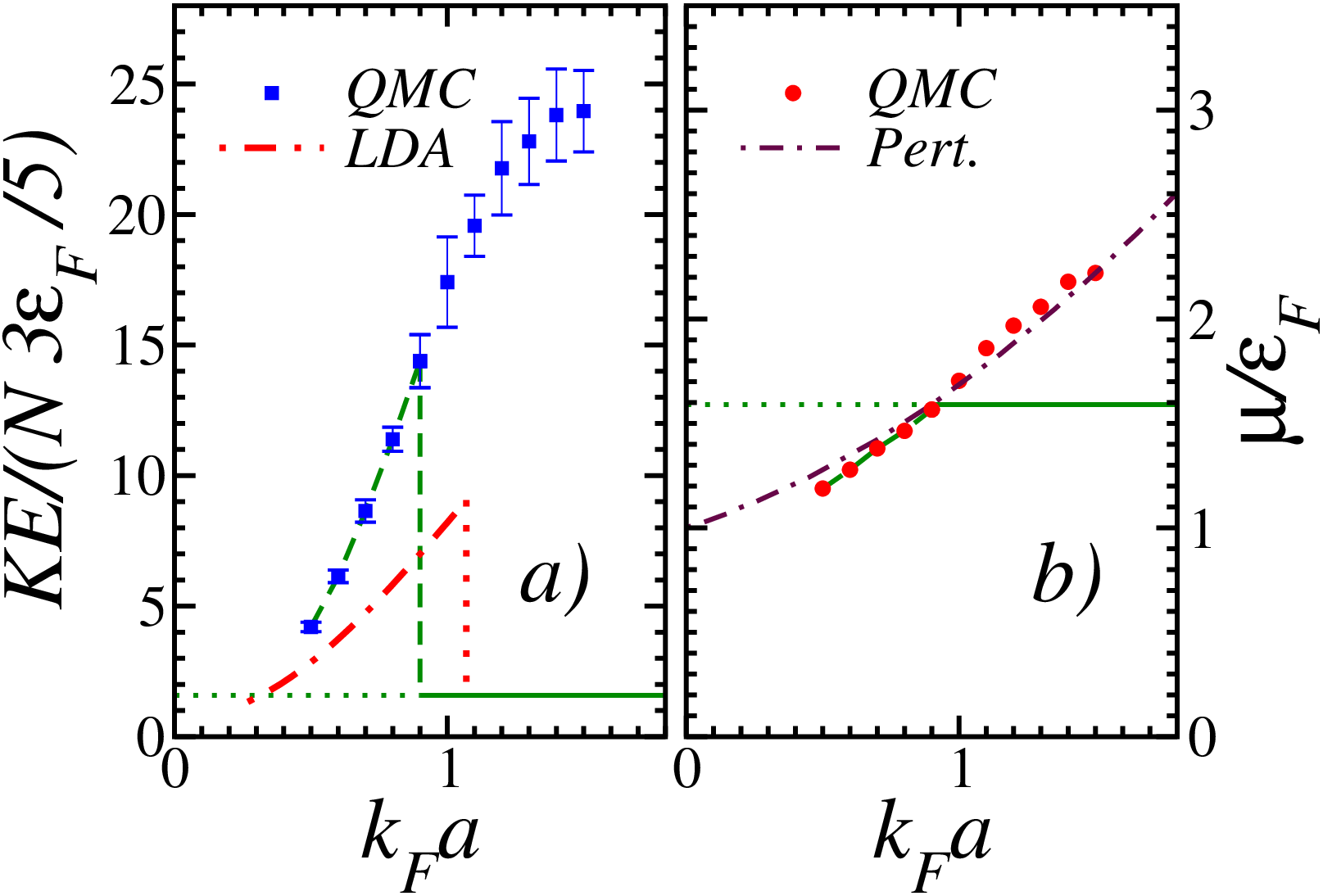}
\caption{(Color online)
(a) Kinetic energy (KE) per particle of the upper branch. Squares represent QMC data
as a function $k_F a$. The dashed line at $k_F a = 0.89$ shows the ferromagnetic transition at which the KE is greatly suppressed and then remains constant. The dot-dash curve is
the LDA result for the KE versus $k_F^0 a$
using the QMC equation of state. Within LDA, ferromagnetism appears at the
center of the trap when $k_F^0 a \simeq 1.1$.
(b) Chemical potential $\mu(k_F a)$, related to the square of the LDA radius
of the trapped cloud, for the upper branch. 
The perturbative result for $\mu = \left(\partial E/\partial N\right)$ 
obtained from Eq.~(\ref{pert-thy}) is also shown.
}
\label{ke-mu}
\end{figure}

In the upper branch, on the other hand, we find
a large cancellation between a positive $\langle K \rangle$ and 
a negative $\langle V \rangle$.
In marked contrast to hard spheres, the upper branch $g(r)$ is extremely large at $r=0$,
has a pronounced dip at the node in the Jastrow $f(r)$ 
and then goes to unity at large $r$; see Fig.~\ref{g2}(a).
For the short-range attraction, the potential energy
$\langle V\rangle \sim g(0) \int d^3{\bf r} V(r)$ is thus large and negative,
dominated by the growth of $g(0)$ with increasing $k_F a$ [inset of Fig.~\ref{g2}(a)]. 
This is compensated by a large positive kinetic energy $\langle K \rangle$
[Fig.~\ref{ke-mu}(a)] so that we find the total energy shown in Fig.~\ref{energy}(a).

\noindent
{\bf Harmonic Trap:}
We first obtain from our QMC data
the chemical potential $\mu = \left(\partial E/\partial N\right)$
as a function of density [Fig.~\ref{ke-mu}(b)].
We then invert this to find the equation of state 
$n(\mu)$ of the \emph{homogeneous} system. 

We restrict ourselves to the paramagnetic Fermi liquid regime here,
and use the LDA $\mu(r) = \mu(0) - V_{\rm trap}(r)$
to study the effects of the harmonic trap
$V_{\rm trap}(r)$ with associated length scale $a_{HO}$.
To compare with experiments, we use
the parameter $k_F^0 = \left( 24 N \right)^{1/6}/ a_{HO}$
as a measure of the total number of particles $N$.
To find the chemical potential at the center 
$\mu(0)$, we solve the LDA equation
$(k_F a)^6  = 2^{3/2}48\pi\int_0^{\tilde{\mu}(0)} d\tilde{\mu}
[\tilde{\mu}(0) - \tilde{\mu}]^{1/2} \tilde{n}(\tilde{\mu})$.
Here we have used dimensionless quantities
$\tilde\mu = {\mu}(0) ma^2$ and
$\tilde{n}(\tilde{\mu}) = n(\mu)a^3$,
where $n(\mu)$ is the QMC equation of state.

We then find the density $n(r=0)$ at the
center of the trap, from which we can determine
the interaction parameter 
$k_F(0) a = [3\pi^2\tilde{n}(\tilde{\mu}(0))]^{1/3}$.
We find that for $k_F^0 a \simeq 1.1$, the trap center 
reaches $k_F(0) a = 0.89$, the critical value in the
homogeneous case. At this point the center of the trap should
become unstable to ferromagnetism.
We have also calculated within LDA the total and kinetic energies
in the trapped system as functions of $k_F^0 a$; the latter
is shown in Fig.~\ref{ke-mu}(a). 

\noindent
{\bf Comparison with experiments:}
While we were motivated by the experiments of Ref.~\cite{ketterle09},
we focus only on ``equilibrium'' in the upper branch, and
do not address dynamical questions. \emph{If} three-body processes leading to
molecule formation can be suppressed, there may be a window of time-scales 
where equilibrium physics in the upper branch, as described here, would be 
observed. The $k_F a$-dependence of $g(r=0)$ [inset of Fig.~\ref{g2}(a)]
is relevant to the loss rate~\cite{petrov03} due to molecule formation.

Even with these caveats, there are some aspects of the experiment 
which we can understand qualitatively and others we cannot.
First, we do find a ferromagnetic instability in the upper branch, but predict that 
it should happen in a homogeneous system at $k_F a = 0.89$, which translates into 
the onset of FM at the center of the trap at $k_F^0 a \simeq 1.1$, while the experiment
sees interesting features only at $k_F^0 a \simeq 2$. The behavior of the chemical potential
[Fig.~\ref{ke-mu}(b)], which increases with increases $k_F a$ and then saturates beyond
the transition is qualitatively consistent with the experiment. However, the 
$k_F a$-dependence of the kinetic energy is not; our results in Fig.~\ref{ke-mu}(a)
are qualitatively different from the experiments. Finally, we have not addressed
here the question of FM domains and their sizes, which is important
to understand given that they have not been seen in the experiment.

\noindent
{\bf Conclusions:}
We show using QMC that fermions with effectively repulsive interactions
become unstable to ferromagnetism beyond a critical interaction strength $k_Fa \simeq 0.9$.
This is true both for fermions in the upper branch (scattering state with positive $a$) of an attractive potential and also for hard sphere repulsion, despite 
important differences in their short range correlations and the kinetic energy that
reflect the underlying potentials. 

\noindent
{\bf Acknowledgments:}
We acknowledge support from ARO W911NF-08-1-0338 and NSF-DMR 0706203 and the use of
computational facilities at the Ohio Supercomputer Center. We thank S.~Zhang for discussions.

\noindent
{\bf Note added:} As we were writing this paper, the work of Pilati {\it et al.}\ appeared
~\cite{pilati}. It addresses the same problem using a similar, 
but not identical, approach. Wherever they overlap, our results are in essential agreement.


\begin{thebibliography}{99}
\bibitem{locv-footnote}
The upper branch solution ($\lambda >0$) for $f(r)$ with a single node
is obtained from the LOCV equation~\cite{vijay1,cowell02}
$\left[ -(1/{m}){d^2}/{dr^2} + V(r) \right]rf(r) = \lambda r f(r)$
for $r \leq$ ``healing length''.
For a zero-range potential 
$f(r) = \sin(\sqrt{m\lambda}(r-b))/r$, with $b$ given by
$\sqrt{m\lambda}b = \tan^{-1}(\sqrt{m\lambda}a)$. 
For $\sqrt{m\lambda}a \rightarrow 0^+$, we thus get
$f(r) \sim (1-a/r)$.
\end{thebibliography}
\end{document}